\documentclass[10pt]{article}

\input epsf

\usepackage{amsmath}
\usepackage{amscd}
\usepackage{amsfonts}

\newcommand{\news}{\setcounter{equation}{0}}
\newcommand{\be}{\begin{equation}}
\newcommand{\ee}{\end{equation}}
\newcommand{\bea}{\begin{eqnarray}}
\newcommand{\eea}{\end{eqnarray}}

\newcommand{\bal}{\begin{align}}
\newcommand{\eal}{\end{align}}

\newcommand{\bean}{\begin{eqnarray*}}
\newcommand{\eean}{\end{eqnarray*}}
\font\upright=cmu10 scaled\magstep1

\newcommand{\Z}{\mathbb{Z}}

\newcommand{\C}{\mathbb{C}}
\newcommand{\R}{\mathbb{R}}
\newcommand{\HH}{\mathbb{H}}
\newcommand{\CP}{\mathbb{CP}}

\newcommand{\identity}{{\upright\rlap{1}\kern 2.0pt 1}}
\newcommand{\half}{\frac{1}{2}}
\newcommand{\pr}{\partial}

\newcommand{\N}{{\cal N}}
\newcommand{\e}{\varepsilon}

\newcommand{\rmd}{{\rm d}}
\newcommand{\rme}{{\rm e}}
\newcommand{\rmi}{{\rm i}}

\begin{document}
\pagestyle{plain}
\title{\vskip -70pt
\begin{flushright}
{\normalsize DAMTP-2009-88} \\
\end{flushright}
\vskip 50pt
{\bf \Large Vortices on Hyperbolic Surfaces}
 \vskip 30pt}
\author{Nicholas S. Manton\thanks{N.S.Manton@damtp.cam.ac.uk} 
\,\,\,and\,\,\,Norman A. Rink\thanks{N.A.Rink@damtp.cam.ac.uk} \\ \\
{\sl Department of Applied Mathematics and Theoretical Physics,}\\
{\sl University of Cambridge,}\\
{\sl Wilberforce Road, Cambridge CB3 0WA, England.}\\
}
\vskip 20pt
\date{March 2010}
\maketitle
\vskip 20pt

\begin{abstract}
It is shown that abelian Higgs vortices on a hyperbolic surface $M$ can be
constructed geometrically from holomorphic maps $f:M \rightarrow N$,
where $N$ is also a hyperbolic surface. The fields depend on $f$ and 
on the metrics of $M$ and $N$. The vortex centres are the ramification 
points, where the derivative of $f$ vanishes. The magnitude of the
Higgs field measures the extent to which $f$ is locally an isometry. 

Witten's construction of vortices on the hyperbolic plane is
rederived, and new examples of vortices on compact surfaces and on 
hyperbolic surfaces of revolution are obtained. The interpretation of 
these solutions as $SO(3)$-invariant, self-dual $SU(2)$ Yang--Mills fields 
on $\R^4$ is also given. 
\end{abstract}

\vskip 80pt

\newpage
\section{Introduction}
\news

In a stimulating paper many years ago, Witten \cite{Witten} constructed all
the $SU(2)$ Yang--Mills instantons on $\R^4$ which are invariant under 
an $SO(3)$ symmetry, $SO(3)$ acting in the standard way on the
$\R^3$ factor in $\R^4 = \R \times \R^3$. Instantons are
solutions of the self-dual Yang--Mills equation for
a pure $SU(2)$ gauge field, and in the first part of the 
paper, Witten showed that the $SO(3)$ symmetry reduces this 
equation to Bogomolny equations for abelian Higgs vortices 
on the hyperbolic plane $\HH^2$. Here the fields are a complex-valued
Higgs field and a $U(1)$ (magnetic) gauge field. In the second half of 
the paper, Witten showed that these Bogomolny equations can in turn 
be reduced to 
Liouville's equation, which can be explicitly solved using a 
holomorphic map $f$ from $\HH^2$ to $\HH^2$. To satisfy boundary 
conditions, and finiteness of the Yang--Mills action, $f$ must be a 
finite Blaschke product. The vortex solution depends both on the 
purely complex information in $f$, and also on the metric on 
$\HH^2$. Each vortex solution, with vortex number $\N$, gives 
an instanton with instanton number $\N$. 

In this way, Witten incidentally 
constructed the complete set of vortex solutions on $\HH^2$, with 
any finite, positive vortex number. The moduli of the vortices are the
vortex centres, the points (counted with multiplicity) 
where the Higgs field vanishes. These points are where the derivative
of $f$ vanishes.

Much is also known about abelian Higgs vortices on flat
$\R^2$, and on compact Riemann surfaces with arbitrary metrics 
\cite{ManSut}. The Bogomolny equations are not integrable, so no 
explicit solutions are known. However, the existence and uniqueness 
of an $\N$-vortex solution on $\R^2$, with the vortex centres at 
$\N$ arbitrarily specified points, was established by 
Taubes \cite{Taubes}. For vortices on a compact surface
$M$, the Higgs field and gauge field are a section and connection on a
(unitary) line bundle $E$ over $M$, and the vortex number $\N$ is the degree of
the bundle $E$. In this setting, Bradlow \cite{Brad} and Garc\'{\i}a-Prada
\cite{Gar} proved a
similar existence result for $\N$-vortices with arbitrary centres, but
subject to the area $A_M$ of $M$ being sufficiently large to accommodate $\N$
vortices. More precisely, $\N$-vortex solutions exist, only if $4\pi \N
< A_M$, this being known as the (strict) Bradlow inequality. The moduli
space of solutions is then ${M^{\N}}_{\rm symm.}$, the $\N$th symmetrized
power of $M$. Solutions also exist when $4\pi\N = A_M$, but they are 
limiting cases of vortices, as the Higgs field vanishes everywhere. 
Vortices can be squeezed on to smaller surfaces, but then they do not
satisfy the Bogomolny equations, and their energy goes up.

In this paper we generalize Witten's approach, in order to
find vortex solutions on surfaces $M$, other than $\HH^2$, that have a 
hyperbolic metric, a metric of constant negative curvature. In 
standard units, the Gauss curvature is required to be
$-\half$. The Gauss--Bonnet theorem tells us that, if $M$ is compact
and has genus $g_M$, and has this curvature, then $M$ has
area $A_M = 4\pi(2g_M - 2)$, so $g_M \ge 2$, and the number of vortices 
allowed by the Bradlow inequality is $\N < 2g_M - 2$. An initial 
investigation of this problem appears in \cite{ChenMan}. The problem's
formal integrability, in a twistorial formulation, has
been demonstrated by Popov \cite{Popov}. 

One of our main results is a reformulation of Witten's construction in a
more general geometrical language. We find that vortices on a
hyperbolic surface $M$ can be expressed in terms of a holomorphic map
$f: M \to N$, where $N$ is also a hyperbolic surface of Gauss curvature
$-\half$. The fields on $M$ depend on the derivative of $f$,
and also on the ratio of the metric on $N$ (pulled back by $f$) to the
metric on $M$. The vortex centres are the ramification points of the 
map $f$, that is, the points where the derivative of $f$ vanishes. 
Using this approach, we have found large classes 
of explicit vortex solutions on some non-compact surfaces. These surfaces 
are well-known quotients of $\HH^2$ by the infinite, discrete group $\Z$.
One example is the once-punctured unit disc, with its
complete hyperbolic metric. The other examples are hyperbolic
cylinders. We have also found special vortex
solutions on certain compact surfaces $M$. These are only semi-explicit, 
as we do not know the relevant metrics explicitly.

These results, though rather special when $M$ is compact, illuminate
the geometrical meaning of vortices on hyperbolic surfaces. At the
complex level, vortices are essentially the same as
ramification points of $f$. However, physically, each vortex extends
over a finite region (heuristically of area $4\pi$). The
extended vortex coincides with a neighbourhood of the ramification
point where the map $f:M \to N$ fails to be close to an isometry. This
is the region where the magnitude of the Higgs field is significantly
less than 1.
 
The structure of this paper is as follows.  In section 2 we review the
Bogomolny equations for vortices on a Riemann surface $M$ with general 
metric. We also review how a vortex solution can be characterised
in terms of a holomorphic section of a holomorphic line bundle $E$ over $M$
and a metric $H$ on (the fibres of) the bundle. In section 3, we present our
main result, showing how vortex solutions on a hyperbolic surface $M$ 
can be obtained from holomorphic maps $f$ from $M$ to another
hyperbolic surface $N$, and we discuss their geometry. 
 
Section 4 is devoted to examples. We first review Witten's solution for 
vortices on $\HH^2$. Then we construct vortex solutions on 
some compact hyperelliptic surfaces. These make use of a 
map to a compact surface of lower genus, and of the (unique) hyperbolic
metrics on the two surfaces. As an example, we find a 4-vortex
solution on a class of hyperelliptic surfaces of genus 5. 
In the third and fourth subsections we find vortex solutions on 
non-compact, hyperbolic surfaces of revolution. Here our results 
are rather explicit. 

In all these cases, the solutions have an interpretation as
$SO(3)$-invariant, self-dual $SU(2)$ Yang--Mills fields on $\R^4$ or discrete
quotients of $\R^4$. Interpreted this way, some are calorons and some
are monopoles. This is discussed in section 5.

\section{Equations for vortices}
\news

Let $M$ be
a Riemann surface with local complex coordinate $z = x + \rmi y$ (and
complex conjugate coordinate ${\bar z} = x - \rmi y$) and a
compatible Riemannian metric with conformal factor $\Omega$,
\begin{align}
 \rmd s^2 = \Omega(x,y) (\rmd x^2 + \rmd y^2) = \Omega(z, {\bar z}) \rmd z \rmd{\bar z} \,.
\end{align}
The fields are locally a complex Higgs field $\phi(x,y)$ and a $U(1)$ 
gauge potential whose components $(a_x, a_y)$ are real. Globally,
there is a $U(1)$ bundle $E$ over $M$, with fibre $\C$, $\phi$ 
is a section of $E$, and $a = a_x dx +a_y dy$ is a connection 1-form. 
It is convenient to use the derivatives 
$\pr_z = \half(\pr_x - \rmi\pr_y)$ and $\pr_{\bar z} 
= \half(\pr_x + \rmi\pr_y)$, and to define $a_z = \half(a_x - \rmi a_y)$
and $a_{\bar z} = \half(a_x + \rmi a_y)$. The magnetic 
field is $B = \pr_x a_y - \pr_y a_x$ and the 2-form field strength
(the curvature of the connection) is $F = \rmd a = B \, \rmd x \wedge
\rmd y$. Equivalently, $F = F_{z{\bar z}} \, \rmd z \wedge \rmd{\bar z}$, where
$F_{z{\bar z}} = \pr_z a_{\bar z} - \pr_{\bar z} a_z = \frac{\rmi}{2} B$.

The first Chern number of the bundle is
\begin{align}
c_1 = \frac{1}{2\pi} \int_{M} F \,.
\end{align}
This is an integer if $M$ is compact, and also if $M$ is non-compact
provided the fields satisfy appropriate boundary conditions. These
boundary conditions are usually that $|\phi| = 1$ and $D_t\phi = 0$ on the
boundary, where $D_t$ denotes the tangential covariant derivative. 

Vortices (for the purposes of this paper) are solutions of the 
coupled Bogomolny equations \cite{Bog,Sam}
\begin{align}
D_{\bar z}\phi &= 0 \,, \\
F_{z{\bar z}}  &= \rmi \, \frac{\Omega}{4}\left(1 - |\phi|^2 \right) \,,
\end{align}
where $D_{\bar z}\phi = \pr_{\bar z}\phi - \rmi a_{\bar z}\phi$ and
$|\phi|^2 = \phi{\bar \phi}$. These equations are gauge covariant
under $U(1)$ gauge transformations, so well-defined for a section and
connection on $E$. They are also covariant under holomorphic changes of 
coordinate, because $F_{z{\bar z}}$ and $\Omega$ transform the same way. The
equations are therefore well-defined globally on the surface $M$. This can be
seen in a more formal way by rewriting the Bogomolny equations as 
equations for forms
\begin{align}
D^{0,1}\phi &= 0\,,\\
F &= \frac{\omega}{2}\left(1 - |\phi|^2 \right) \,,
\end{align}
where $\omega=\frac{\rmi}{2}\,\Omega\,\rmd z \wedge \rmd{\bar z}$ is the
K\"ahler 2-form on $M$ and $D^{0,1}$ denotes
the (0,1)-part of the covariant derivative which is a notion invariant under 
holomorphic coordinate changes. Non-trivial solutions of these
Bogomolny equations exist only if the Chern number is positive.

The vortex centres are the points on $M$ where $\phi$ vanishes, and
the number of vortices $\N$ (counted with multiplicity) equals the Chern
number. Moreover, the first Bogomolny equation ensures that the
multiplicities are all positive.

There is a mathematical reformulation (c.f. \cite{Chern}) of the 
Bogomolny equations, which we now describe. Introduce a bundle metric 
$H$, locally 
a positive real function on $M$, and replace $|\phi|^2$ by $|\phi|^2_H 
= H\phi{\bar \phi}$ in the second Bogomolny equation. In the unitary 
framework we have used so far, $H = 1$ for any choice of (unitary)
gauge. However, the Bogomolny equations now have a larger gauge 
freedom. We may perform a gauge transformation $g(z,{\bar z})$ with 
values in $\C^{\, *}$, the non-zero complex numbers, having the effect
\begin{align}
 &\phi \rightarrow g\phi \,, \\
 &{\bar \phi} \rightarrow {\bar g}{\bar \phi} \,, \\
 &a_z \rightarrow a_z - \rmi (\pr_z g)\,g^{-1} \,,  
	\quad a_{\bar z} \rightarrow 
      a_{\bar z} - \rmi (\pr_{\bar z}  g)\,g^{-1} \,, \\
 &H \rightarrow g^{-1}{\bar g}^{-1} H \,. 
\end{align}
The magnitude of the Higgs field, $|\phi|^2_H$, is gauge invariant in
this larger sense, and so is $F_{z{\bar z}}$. Note that $\bar\phi$ is 
a section of the bundle $\bar E$, and the metric $H$ can be regarded 
as a section of ${\bar E}^* \otimes E^*$, the tensor product of the
dual bundles. Therefore the covariant derivatives of $\phi$, $\bar \phi$
 and $H$ are
\begin{align}
 &D_z \phi = \pr_z \phi - \rmi a_z \phi \,, \\
 &D_z {\bar \phi} = \pr_z {\bar\phi} 
+ \rmi\overline{a_{\bar z}}{\bar \phi} \,, \\
 &D_z H = \pr_z H + \rmi a_z H - \rmi \overline{a_{\bar z}}H \,,
\end{align}
and similarly for $D_{\bar z}$. In the unitary gauge 
\begin{align}
D_z H = 0 \,,
\end{align}
so this covariant derivative vanishes in any gauge. (Also $D_{\bar z} H
= 0$ by hermitian conjugation.)

In this formulation, the Bogomolny equations become the coupled system
\begin{align}
D_{\bar z}\phi &= 0 \,, \\
D_z H &= 0 \,, \\
F_{z{\bar z}} &= \rmi \, \frac{\Omega}{4}\left(1 - H\phi{\bar \phi} \right) \,.
\end{align}

One can still work in the unitary gauge, with $H = 1$, but now one may
also transform to a holomorphic gauge. This is where $a_{\bar z} =
0$ everywhere. By Dolbeault's Lemma (see e.g. \cite{Forster}) such 
a gauge always exists. 
The gauge transition functions between trivializations of $E$ on
overlapping patches of $M$ must now be holomorphic, and therefore the
bundle $E$ is a holomorphic line bundle.
Furthermore, the first Bogomolny equation reduces to 
\begin{align}
\pr_{\bar z} \phi = 0 \,,
\end{align}
so $\phi$ is a holomorphic section of $E$, locally written
$\phi(z)$. The value of $\phi$ at a point is not gauge invariant,
because of the remaining freedom to perform holomorphic gauge
transformations $g(z)$, but the positions of the zeros of $\phi$ are
(as before) gauge invariant, and define the vortex centres.

In holomorphic gauge, the second Bogomolny equation becomes 
\begin{align}
\pr_z H + \rmi a_z H = 0 \,,
\end{align}
so 
\begin{align}
a_z = \rmi\pr_z(\log H) \,, \quad a_{\bar z} = 0 \,,
\end{align}
which is called the Chern connection. We see that $H$ is a fundamental 
quantity in holomorphic gauge. The curvature of the Chern connection is
\begin{align}
F_{z{\bar z}} = - \pr_{\bar z}a_z = -\rmi\pr_z\pr_{\bar z}(\log H) \,,
\end{align}
so the final Bogomolny equation reduces to
\begin{align}
\pr_z\pr_{\bar z}(\log H) = -\frac{\Omega}{4}\left(1 - 
H\phi(z)\overline{\phi(z)} \right) \,,
\label{RedBogo2}
\end{align}
with $\phi$ holomorphic. This is the key equation that remains to be
solved. It is a version of Taubes' equation \cite{Taubes}. 

\section{Hyperbolic vortices from holomorphic maps}
\news

For a general surface with metric $\rmd s^2 = 
\Omega(z,{\bar z})\rmd z \rmd{\bar z}$,
 the Gauss curvature is 
\begin{align}
K = -\frac{2}{\Omega}\pr_z\pr_{\bar z}(\log\Omega) \,.
\end{align}
The surface has constant curvature $-\half$ if
\begin{align}
\pr_z\pr_{\bar z}(\log\Omega) = \frac{\Omega}{4} \,,
\label{Liou}
\end{align}
which is Liouville's equation. The surface and its metric are then
called hyperbolic.

Let $M$ and $N$ be Riemann
surfaces, carrying hyperbolic metrics $\Omega_M$ and $\Omega_N$ of
curvature $-\half$. Let $f:M \rightarrow N$ be a non-constant holomorphic
mapping. In terms of a complex coordinate $z$ on some neighbourhood
$U$ of $M$, and coordinate $w$ on its image $f(U)$, the map is defined
by a holomorphic function $w=f(z)$.

Using the map $f$, we construct a vortex solution on $M$ as follows. 
Working in holomorphic gauge, 
we set $\phi(z) = \frac{\rmd f}{\rmd z}$. $\phi$
vanishes at a discrete set of points, the ramification points of the
map, and these are the vortex centres. Locally, around a ramification
point and its image, one may find new coordinates ${\tilde z}, 
{\tilde w}$ so that ${\tilde w} = {\tilde z}^k$ for some integer 
$k$ greater than 1. The ramification index is $k-1$, 
and this is the multiplicity of the vortex.

It remains to determine $H$, so that it satisfies the Bogomolny
equation
\begin{align}
\pr_z\pr_{\bar z}(\log H) = -\frac{\Omega_M}{4}\left(1 - H\phi(z)\overline{
  \phi(z)} \right) \,.
\end{align}
Set $H = H_1/H_2$, so
\begin{align}
\pr_z\pr_{\bar z}(\log H_1) - \pr_z\pr_{\bar z}(\log H_2)
= -\frac{\Omega_M}{4} + \frac{\Omega_M}{4}\frac{H_1}{H_2}\phi{\bar \phi} \,.
\label{H1H2split}
\end{align}
Choosing $H_2 = \Omega_M$, which satisfies (\ref{Liou}),
equation~(\ref{H1H2split}) simplifies to
\begin{align}
\pr_z\pr_{\bar z}(\log H_1) = \frac{1}{4}H_1\phi{\bar \phi} \,.
\label{H1eq}
\end{align}
Now $\Omega_N(w,{\bar w})$ satisfies the Liouville equation (on $N$)
\begin{align}
\pr_w\pr_{\bar w}(\log\Omega_N) = \frac{1}{4}\Omega_N \,,
\end{align}
so its pullback to $M$, defined by $f^*\Omega_N(z,{\bar z}) = 
\Omega_N(f(z),\overline{f(z)})$, satisfies
\begin{align}
\pr_z\pr_{\bar z}(\log f^*\Omega_N) =
\frac{1}{4}f^*\Omega_N \frac{\rmd f}{\rmd z}
\overline{\frac{\rmd f}{\rmd z}} \,,
\end{align}
by the chain rule. Therefore, (\ref{H1eq}) is solved by setting
$H_1 = f^*\Omega_N$, since $\phi = \frac{\rmd f}{\rmd z}$. In summary, 
given the map
$f:M \rightarrow N$, a vortex solution on $M$ is obtained by setting
\begin{align}
\phi = \frac{\rmd f}{\rmd z} \quad {\rm and} \quad
 H = \frac{f^*\Omega_N}{\Omega_M} \,,
\label{phiandH}
\end{align}
and hence
\begin{align}
|\phi|^2_H = \frac{f^*\Omega_N}{\Omega_M}
\frac{\rmd f}{\rmd z}\overline{\frac{\rmd f}{\rmd z}} \,.
\label{HiggsfinalMN}
\end{align}

This description is local, but we can make it global. The construction
itself tells us which line bundle the Higgs field $\phi$ is a
section of. Recall that if $f$ maps $M$ holomorphically to $N$, then
$\rmd f$ (the derivative) maps $TM$ to $TN$, where these are the
holomorphic tangent bundles. Equivalently, $df$ is a section of the
line bundle $E = TM^* \otimes f^*TN$ over $M$, where the second factor
is the pulled-back tangent bundle of $N$, and $TM^*$ is the dual of
the tangent bundle of $M$. Therefore the Higgs
field $\phi$ is a section of this line bundle $E$. The formula
for $H$ in (\ref{phiandH}) is clearly (indeed
canonically) a metric on $E$.

An elegant way to see that $|\phi|^2_H$ is a globally defined function
on $M$ is to identify it as the ratio of 2-forms. The K\"ahler forms on $M$
and $N$ are
\begin{align}
\omega_M &= \frac{\rmi}{2}\Omega_M(z,\bar z) \rmd z\wedge \rmd{\bar z} \,, \\
\omega_N &= \frac{\rmi}{2}\Omega_N(w,\bar w) \rmd w\wedge \rmd{\bar w} \,,
\end{align}
and the map $f$ gives the pull-back $f^*\omega_N$ on $M$.
One verifies in local coordinates that (\ref{HiggsfinalMN})
agrees with
\begin{align}
|\phi|^2_H = \frac{f^*\omega_N}{\omega_M} \,. 
\label{eq:Higgs_ratio}
\end{align}

We easily obtain a global understanding of the degree of the bundle
$E$ when $M$ and $N$ are compact (in which case $f$ is surjective).
$TM^*$ is a line bundle of degree $2g_M - 2$, and $TN$ (over $N$) is of
degree $-(2g_N - 2)$. Let the degree of the mapping $f$ be $n$; i.e. away
from the images of ramification points, each point of $N$ has $n$
preimages. Then $E$ has degree (Chern number)
\be
 2g_M - 2 - n(2g_N - 2) \,.
\label{Edeg}
\ee
By the Riemann--Hurwitz formula \cite{FarKra}, this is precisely the total
ramification number of $f$ (the sum of the ramification indices at all
the ramification points). The vortex number ${\cal N}$ is the
Chern number of the bundle $E$, and hence equal to the total ramification 
number of $f$, as expected. 

More precisely, the set of vortex centres (with their
multiplicities) coincides with the ramification divisor of $f$, and
this defines the divisor class of $E$. The ramification divisor
class is the canonical class on $M$ minus the pullback of the
canonical class on $N$, but we will not use this observation here.

This discussion shows that the vortex solutions we can
construct on a compact surface $M$ using maps $f$ are rather 
special. From (\ref{Edeg}) it is
obvious that the vortex number must be even. To have ramification 
points at all, $n$ must be 2 or more, and the expression (\ref{Edeg})
must be positive. There are rather few maps from a given surface $M$, with 
$g_M \ge 2$, to another
surface $N$, with $g_N \ge 2$, so the possible ramification divisors 
on $M$ are limited. The smallest genus allowing a nontrivial 
solution is $g_M = 4$. Then with $n=2$ and $g_N = 2$, a 2-vortex 
solution is possible. We shall describe an explicit example in
subsection 4.2.

Note that (\ref{Edeg}) implies that the vortex number ${\cal N}$
is strictly less than $2g_M - 2$. From the Gauss--Bonnet theorem, 
and given that $M$ has curvature $-\half$, the area of $M$ is 
$A_M = 4\pi(2g_M - 2)$. So our vortex solutions, as expected, satisfy 
the Bradlow inequality $4\pi{\cal N} < A_M$.

We conclude with some remarks on the geometrical interpretation of
these hyperbolic vortices. Note first that since $M$ and $N$ are
hyperbolic, with the same curvature, locally there are isometries
between them. However, $f: M \rightarrow N$ is not globally, nor
locally, an isometry. A 
necessary condition for a degree $n$ map to be an isometry is that 
$A_M = nA_N$. But expression (\ref{Edeg}) implies that $A_M - nA_N$ is
$4\pi$ times the vortex number ${\cal N}$. So the number of vortices
is a global measure of the failure of $f$ to be an isometry. More 
precisely, this failure is captured by the expression (\ref{eq:Higgs_ratio}) 
for $|\phi|^2_H$, whose value determines the extent to
which $f$ is locally an isometry. Where $|\phi|^2_H$ is (close to) 1, $f$ is
(close to being) an isometry. However, $|\phi|^2_H$ is zero at the
vortex centres, and small nearby. So vortices occupy those regions of 
$M$, centred at the ramification points of $f$, where $f$ is not close 
to being an isometry. The regions are not precisely defined, but 
heuristically, each simple vortex has an area $4\pi$. Hyperbolic vortices can
still be interpreted in the traditional way as topological solitons
carrying magnetic flux, but the preceding discussion shows that they 
can also be interpreted purely geometrically. 

There is an interesting analogy with Skyrmions here. The energy
excess of a Skyrmion above the Faddeev--Bogomolny lower bound is 
a measure of the failure of the Skyrme field to be an isometry \cite{ManSut}. 

\section{Examples}
\news

\subsection{Vortices on the hyperbolic plane}

In the upper half plane (UHP) model, $\HH^2$ is represented by the
region ${\rm Im} \, z > 0$, and the metric is
\begin{align}
 \rmd s^2 = \frac{2}{({\rm Im} \, z)^2} \rmd z \rmd{\bar z} \,,
\label{Hypmetric}
\end{align}
satisfying (\ref{Liou}). Vortices on $\HH^2$ are obtained using maps 
$f:\HH^2 \to \HH^2$. Let the target $\HH^2$ also be represented by the 
UHP, with complex coordinate $w$ and metric 
$\rmd s^2 = 2 ({\rm Im} \, w)^{-2} \rmd w \rmd{\bar w}$.
A map $f$ is now simply a function $w = f(z)$.

Using the results of the last section, we see that a vortex solution 
can be obtained by setting 
\begin{align}
\phi = \frac{\rmd f}{\rmd z} \quad {\rm and} \quad
 H = \frac{({\rm Im} \, z)^2}{({\rm Im} f(z))^2} \,,
\end{align}
and hence the magnitude of the Higgs field is
\begin{align}
|\phi|^2_H = \frac{({\rm Im} \, z)^2}{({\rm Im} f(z))^2}
\frac{\rmd f}{\rmd z}\overline{\frac{\rmd f}{\rmd z}} \,.
\label{Higgsfinal}
\end{align}
The connection is the Chern connection,
\begin{align}
a_z = \frac{1}{{\rm Im} \, z} - \frac{1}{{\rm Im} f}\frac{\rmd f}{\rmd z} \,,
\quad a_{\bar z} = 0 \,.
\end{align}

To satisfy the boundary condition $|\phi|^2_H = 1$ when ${\rm Im} \, z = 0$,
and to have a finite vortex number $\N$, $f$ must be a Blaschke 
function
\begin{align}
f(z) = -\rmi \, \frac{\prod^{\N}_{i=0}(z - a_i) 
- \prod^{\N}_{i=0}(z - \overline{a_i})}
{\prod^{\N}_{i=0}(z - a_i) + \prod^{\N}_{i=0}(z - \overline{a_i})}
\label{Blaschke}
\end{align}
with ${\rm Im} \, a_i > 0 \,, \forall i$. Geometrically, $f$ is a holomorphic
mapping of the UHP to itself, also mapping the boundary to
itself~\footnote{A standard Blaschke function is $f(z) =
\prod^{\N}_{i=0} \left(\frac{z - a_i}{z - \overline{a_i}} \right)$,
which maps the UHP to the unit disc, and (\ref{Blaschke}) is a simple M\"obius
transformation of this.}. The topological degree of the mapping is
$\N +1$. $\frac{\rmd f}{\rmd z}$ vanishes at $2\N$ points, but 
$\N$ of these are
outside the UHP and can be disregarded. The remaining $\N$ points are
the ramification points of $f$ in the UHP. These are the vortex centres,
where $|\phi|^2_H = 0$. Their positions in $\HH^2$ are arbitrary.

In summary, this construction (due to Witten) produces the complete
set of $\N$-vortex solutions on $\HH^2$ in terms of degree $\N +1$
holomorphic maps from $\HH^2$ to $\HH^2$, and using the hyperbolic metrics on
both domain and target (as one sees from formula
(\ref{Higgsfinal})). The vortex centres are the ramification points. 

\subsection{$M$ compact}

There are numerous examples of holomorphic maps between compact Riemann 
surfaces $M$ and $N$, both of genus 2 or more. However, by a theorem
of de Franchis, these maps are isolated, and do not have moduli
\cite{Kob}. So the vortex solutions we obtain, using formulae 
(\ref{phiandH}), are rather special. These solutions are also
only semi-explicit; although we can specify the map $f:M \rightarrow
N$, we do not have explicit expressions here for the hyperbolic metrics on $M$
and $N$. Such explicit expressions for the metrics would be available
if $M$ and $N$ were given as quotients of $\HH^2$ by suitable Fuchsian groups.

A simple class of examples is as follows. Let
$M$ be a hyperelliptic surface defined by
\bea
\eta^2 &=& (z - e_1)(z + e_1) \dots (z - e_r)(z + e_r) \\
    &=& (z^2 - e_1^2) \dots (z^2 - e_r^2)
\eea
with $r \ge 5$, and all $\pm e_i$ distinct and non-zero. This is a
compact double covering of ${\CP}^1$. $M$ has two sheets over
a neighbourhood of $z=0$, and two sheets over a neighbourhood of $z =
\infty$. In addition to the hyperelliptic involution 
$J: z \to z \,, \, \eta \to -\eta$ 
which exchanges sheets, there is a further involution $I:z \to -z
\,, \, \eta \to \eta$. If we quotient by the involution $I$ we obtain the 
hyperelliptic surface $N$ defined by
\begin{align}
\eta^2 = (w - e_1^2) \dots (w - e_r^2) \,.
\end{align}
$N$ has two sheets over a neighbourhood of $w=0$.
Over a neighbourhood of $w = \infty$, $N$ has only one sheet if $r$ 
is odd, but two sheets if $r$ is even.
The (projection) map $f$ from $M$ to $N$ is given by $(z,\eta) \to
(w,\eta)$, where $w = f(z) = z^2$,
so $\frac{\rmd f}{\rmd z} = 2z$.
  
The genus of $M$ is $g_M = r-1$, and the genus of $N$ is $g_N = \half(r-1)$
if $r$ is odd and $\half(r-2)$ if $r$ is even. The map $f$ has
ramification points of index 1 at $z=0$ (one on each sheet). If $r$ is
even, then additionally, there are ramification points of index 1 at 
$z=\infty$ (one on each sheet). Our construction of vortex solutions,
using $f$, therefore gives a 2-vortex on $M$ if $r$ is odd, and a
4-vortex if $r$ is even. The vortex count is confirmed using the
Riemann--Hurwitz formula. The lowest genus examples are for $r=5$ and $r=6$,
where $g_M = 4$ and $g_M = 5$, respectively, and $g_N=2$.

These hyperelliptic surfaces have unique hyperbolic metrics with
finite area, and in terms of these metrics, the fields $H$, $|\phi|^2_H$ and
$a_z$ could be found. Unfortunately, we cannot be more explicit.

This kind of example generalizes. Given any compact Riemann surface $M$ with
a non-trivial, conformal automorphism group $G$ (the full automorphism group
can be bigger than this, with $G$ a subgroup of it), we can quotient by
$G$. The quotient is a Riemann surface $N$, and we define the map $f$
to be the natural projection. $f$ has degree $|G|$. The
ramification points of $f$ are the points of $M$ fixed by any subgroup
of $G$ with more than one element. Provided $M$ and $N$
are both of genus 2 or more, then they have hyperbolic metrics, and we
can construct a vortex solution on $M$ with vortex centres at the
ramification points of $f$. Since by the Schwarz and Hurwitz theorems
(see e.g. \cite{FarKra}) any compact surface $M$ has a finite automorphism
group, these vortex solutions are again discrete, and do not have moduli.

\subsection{$M$ a once-punctured disc}

In this subsection and the next we will construct explicit families of vortex
solutions on the hyperbolic Riemann surfaces which are obtained by
quotienting $\HH^2$ by a $\Z$-action. As complex manifolds, one of these 
surfaces is the open unit disc with a puncture at the origin. The rest are
open cylinders, parametrised by one real conformal
invariant. These surfaces acquire geodesically complete hyperbolic
metrics from $\HH^2$, and become surfaces of revolution (parts of 
which can be embedded in $\R^3$). For sketches of these surfaces, 
see e.g. \cite{HilCoh}.

We briefly comment on the boundary conditions that must be imposed:
As one moves out to spatial infinity in $\R^2$ or 
$\HH^2$, the usual boundary condition for vortex solutions is 
$|\phi|_H^2 \to 1$. This is to ensure that the magnetic field decays 
to zero, and the vortex solutions have finite energy. More generally, 
if $M$ is non-compact, and (part of) its boundary is a hyperbolic 
end, analogous to the boundary of $\HH^2$, we again require 
$|\phi|_H^2\to 1$. This can be achieved by a map $f: M \rightarrow N$ 
provided $N$ also has a hyperbolic end, and $f$ maps the
boundary of $M$ to the boundary of $N$. To see this, we use convenient
coordinates, so that both $M$ and $N$ look locally like the UHP and $f$ 
is real on the real axis of $M$, with a non-zero derivative.
At $z_0 \in \R$ let $f(z_0) = f_0 \in \R$, and let $\frac{\rmd f}{\rmd z} =
c_0 \in \R$, with $c_0 \ne 0$. Then, expanding around $z_0$, 
$f(z) = f_0 + c_0(z-z_0) + \cdots$, so ${\rm Im} \, z = 
{\rm Im} \, (z-z_0)$ and ${\rm Im} \, f(z) = c_0 \, {\rm Im} \, (z-z_0) +
\cdots$, and eq.~(\ref{Higgsfinal}) implies that
$|\phi|_H^2(z_0,\overline{z_0}) = 1$.

In the remainder of this subsection we consider the surface $D^*$, 
obtained from the UHP model of $\HH^2$ by
identifying points under the translation $z \to z + 2\pi$, which
generates a $\Z$-action. $D^*$ is therefore the strip 
$\{0 \le {\rm Re} \, z \le 2\pi \,, \, {\rm Im} \, z > 0\}$, with its
edges ${\rm Re} \, z = 0$ and ${\rm Re} \, z = 2\pi$ identified. 
The metric acquired from $\HH^2$ is the metric (\ref{Hypmetric}).

A more convenient coordinate on $D^*$ is $u = \rme^{iz}$, with the range
$0 < |u| < 1$, which covers $D^*$ once. This shows that $D^*$ is an
open disc, punctured at the origin. In terms of $u$, the metric is
\begin{align}
 \rmd s^2 = \frac{8}{u\bar{u}\left(\log u\bar{u}\right)^2} 
 \rmd u \rmd{\bar{u}} \,. 
\label{punctmetric}
\end{align}
So $D^*$ is a surface of revolution, with curvature
$-\half$. The boundary at $|u| = 1$ is a hyperbolic end, but that at
$u=0$ is not. The point $u=0$ is infinitely far from any point of
$D^*$, but the area of the region $0 < |u| < \e$ (with $\e < 1$) is
finite. We refer to the neighbourhood of $u=0$ as a parabolic end.

We first consider a holomorphic map $f: D^* \to D^*$, with the metrics
on both domain and target as described above. The map is clearly
bounded ($|f|<1$) and can therefore be extended to a holomorphic map 
on the complete disc $D=\{ u\in\C \colon |u|<1 \}$, but the maps on 
$D$ with boundary behaviour $|f(u)|\to 1$ as $|u|\to 1$ are precisely 
the Blaschke functions
\begin{align}
f(u) = \prod^{\N}_{i=0}\frac{u - a_i}{1 - \overline{a_i} u} \,,
\label{eq:Blaschke_D}
\end{align}
with $a_i \in D$. Since $f(D^*) \subset D^*$, $f(u)$ cannot vanish for 
any $u \ne 0$, and therefore all $a_i$ must be zero. Thus $f(u)=u^{\N+1}$, and
\begin{align}
|\phi|^2_H = \frac{u\bar{u}\left(\log u\bar{u}\right)^2}
{(u\bar{u})^{\N+1}\left(\N+1\right)^2\left(\log u\bar{u}\right)^2} 
(\N+1)^2(u\bar{u})^{\N} = 1 \,. 
\end{align}
For any $\N$, this is the vacuum solution on $D^*$, with zero vortex 
number. The result also follows from the observation that 
$u \to u^{\N+1}$ is an isometry.

More interesting solutions can be obtained using maps $f: D^* \to D$, where
$D=\{ w\in\C \colon |w|<1 \}$ is the Poincar\'e disc model of $\HH^2$ 
with metric
\begin{align}
 \rmd s^2 = \frac{8}{(1-w\bar w)^2} \rmd w \, \rmd{\bar w} \,,
\end{align}
and curvature $-\half$. By the same reasoning as above, the domain of
$f$ can be extended to $D$ and thus the map is again a
Blaschke function $w = f(u)$ as in (\ref{eq:Blaschke_D}). Using this 
to construct a vortex solution, we find that the Higgs field magnitude 
is
\begin{align}
|\phi|^2_H(u,\bar u) = \frac{u\bar{u} (\log u\bar{u})^2}
{\left(1 - f(u)\overline{f(u)}\right)^2} \frac{\rmd f}{\rmd u} 
\overline{\frac{\rmd f}{\rmd u}} \,. 
\label{eq:cusp_sol}
\end{align}
The derivative $\frac{\rmd f}{\rmd u}$ has precisely $\N$ zeros inside 
the unit disk $D$, and these are the moduli of the vortex solution. 

The Blaschke product (\ref{eq:Blaschke_D}) depends 
on $\N+1$ parameters $\{a_i\}$, but it was observed by Witten 
\cite{Witten} that this is a redundancy in the description of the vortices. 
Following Strachan \cite{Strac}, we remove the redundancy by 
considering only Blaschke products of the form
\begin{align}
f(u) = u \prod^{\N}_{i=1}\frac{u - a_i}{1 - \overline{a_i}u} \,.
\label{BlaschProd}
\end{align}

The simplest such function, $f(u)=u$, leads to the solution
\begin{align}
|\phi|^2_H(u,\bar u) = \frac{u\bar{u} (\log u\bar{u})^2}
{\left(1 - u\bar u\right)^2} \,,
\label{eq:novortex}
\end{align}
which is non-zero in $D^*$. As expected at a hyperbolic end, 
$|\phi|^2_H(u,\bar u) \to 1$ as $|u|\to 1$. By contrast, at the
parabolic end,
\begin{align}
|\phi|^2_H(u,\bar u) \begin{CD}@>\phantom{u\to 0}>{u\to 0}>\end{CD} 0.
\end{align}
However, the solution should not be thought of as 
a true vortex centred at $u=0$. Firstly, the vortex cannot move as there 
are no moduli; secondly, the point $u=0$ is outside $D^*$ and infinitely
far away. 

The calculation of the first Chern number of this solution yields $c_1=1$:
\begin{align} 
 c_1 &= \frac{1}{2\pi}   \int_{D^*} F \\
     &= -\frac{\rmi}{2\pi} \int_{D^*} \pr_{u}\pr_{\bar u} (\log H) 
        \, \rmd u \wedge \rmd\bar u \\
     &= -\frac{\rmi}{2\pi} \oint_{|u|\to 1} \pr_{\bar u} (\log H)\, \rmd\bar u 
        +\frac{\rmi}{2\pi} \oint_{|u|\to 0} 
         \pr_{\bar u} (\log H) \, \rmd\bar u \,, 
\end{align}
by Stokes' Theorem. Since $\frac{\rmd f}{\rmd u} = 1$ here, 
$H(u,\bar u) = |\phi|^2_H(u,\bar u)$. Then, expressing (\ref{eq:novortex}) 
in polar coordinates, $u = \rho \, \rme^{\rmi\theta}$, the contour 
integrals can be rewritten as
\begin{align} 
 c_1 
 &= -\lim_{\rho\to 1} \frac{1}{2\pi} \int_{0}^{2\pi} 
    \left(\frac{1+\rho^2}{1-\rho^2} + \frac{1}{\log\rho} \right) 
    \rmd\theta \nonumber \\
 &\phantom{=}\, + \lim_{\rho\to 0} \frac{1}{2\pi} \int_{0}^{2\pi} 
    \left(\frac{1+\rho^2}{1-\rho^2} + \frac{1}{\log\rho} \right) 
    \rmd\theta \label{c1} \\
 &= 0 + 1 \,. 
\end{align}
The magnetic field and energy are concentrated around the parabolic end, 
and the total magnetic flux is that of a simple vortex. 

We can add true vortices to this basic configuration, for example by taking
\begin{align}
f(u) = u \, \frac{u - a}{1 - \bar a u} \,.
\label{1addvort}
\end{align}
Then,
\begin{align}
|\phi|^2_H = \frac{ u\bar{u} (\log u\bar{u})^2 (a-2u+\bar{a}u^2) 
(\bar{a}-2\bar{u}+a\bar{u}^2) }
 { (1-u\bar{u})^2 (1-\bar{a}u-a\bar{u}+u\bar{u})^2 } \,. 
\end{align}
This has unchanged boundary behaviour, $|\phi|^2_H\to 1$ as $|u|\to 1$, and
$|\phi|^2_H\to 0$ as $u \to 0$. $|\phi|^2_H$ has a single zero 
inside $D^*$ at
\begin{align}
U = \frac{1-\sqrt{1-a\bar{a}}}{\bar{a}} \,.
\end{align}
Conversely, to add a simple vortex at $U$, choose
\begin{align}
a = \frac{2U}{1+U\bar{U}}
\end{align}
in (\ref{1addvort}). In terms of $u$ and $U$ (and their complex 
conjugates),
\begin{align}
|\phi|^2_H = \frac{4 u\bar{u} (\log u\bar{u})^2 (u-U)(\bar{u}-\bar{U}) 
(\bar{U}u-1)(U\bar{u}-1)} 
{(1-u\bar{u})^2 ( (1+U\bar{U})(1+u\bar{u}) - 2\bar{U}u - 2U\bar{u})^2} \,.
\end{align}

With $f$ of degree 3,
\begin{align}
f(u) = u \, \frac{u - a}{1 - \bar a u} \frac{u - b}{1 - \bar b u} \,, 
\label{doublevortex}
\end{align}
there are two vortices inside $D^*$. Their positions are
the zeros of a polynomial of degree 4, whose cumbersome solution is 
unlikely to give insight into the vortex fields.

We remark that $|\phi|^2_H \to 0$ as $u \to 0$ for any solution of the form
(\ref{eq:cusp_sol}) constructed with a Blaschke function 
of type (\ref{BlaschProd}). The solution 
(\ref{eq:novortex}) has $c_1 = 1$, the lowest Chern number in
this whole sector of vortex solutions on $D^*$, which justifies the 
notion that it is a meta-vacuum. It can be shown that the general
solution in this sector, which has $\N$ true vortices, has Chern 
number $\N + 1$.

\subsection{$M$ a hyperbolic cylinder}

The hyperbolic cylinder, $M$, is defined as follows. Start with the UHP
model of $\HH^2$ with metric (\ref{Hypmetric}). The map $z \to
\rme^{-\lambda}z$, with $\lambda$ real and positive, is an isometry,
and generates an isometric $\Z$-action. $M$ is the quotient space, which
we take to be the half-annulus $\{\rme^{-\lambda} \le |z| \le 1 \,, 
\, {\rm Im} \, z > 0\}$, with the semicircular edges
identified. The metric on the UHP descends to a hyperbolic metric on
$M$. This metric is complete, and has two hyperbolic ends where $M$
approaches the real $z$-axis. 
The surface is characterised conformally by $\lambda$.

Now set $u = - \rmi \log z$. On $M$, the range of $u$ is the rectangle 
$\{0 < {\rm Re} \, u < \pi \,, \, 0 \le {\rm Im} \, u \le \lambda\}$, with 
the opposite edges ${\rm Im} \, u = 0$ and ${\rm Im} \, u = \lambda$ 
identified. The hyperbolic metric, in terms of $u$, is
\begin{align}
 \rmd s^2 = \frac{2}{(\sin({\rm Re} \, u))^2} \rmd u \rmd{\bar u} \,.
\end{align}
From this we see that $M$ is a surface of revolution, with lines at
fixed ${\rm Im} \, u$ being geodesics, and translations in the 
${\rm Im} \, u$ direction being isometries. The circles at fixed 
${\rm Re} \, u$ are generally not geodesics, except for the shortest circle at 
${\rm Re} \, u = \frac{\pi}{2}$, which has length $\sqrt{2}\lambda$.
$M$ has hyperbolic ends at ${\rm Re} \, u = 0$ and ${\rm Re} \, u = \pi$.

To construct vortices on $M$, we use a map $f:M \rightarrow
\HH^2$. The simplest such map is the scaled Jacobi elliptic function
\cite{WhiWat}
\begin{align}
w = f(u) = {\rm sn}\left(\frac{2K}{\pi}u \,; k \right) \,.
\end{align}
This function has real period $2\pi$ 
and imaginary period $\rmi\pi\frac{K'}{K}$, where $K(k),K'(k)$ are the usual
complete elliptic integrals. Given $\lambda$, we choose the unique
value of $k$ in the range $0 < k < 1$, such that
$\pi\frac{K'(k)}{K(k)} = \lambda$.
The Jacobi function maps a period rectangle on to the extended complex plane,
but it maps $M$ (which occupies half a period rectangle) on to
the right hand half-plane only. This can be verified by using the known values 
of the function on the boundary of $M$ (deformed by small semicircles 
around the poles), and using the property that the real part of the function 
is harmonic, so that its maximal and minimal values occur on the
boundary. Metrically, we identify the image of $f$ to be $\HH^2$ in the 
right hand half-plane model. Using the coordinate $w$, the metric on
the half plane is 
\begin{align}
\rmd s^2 = \frac{2}{({\rm Re} \, w)^2} \rmd w \rmd{\bar w} \,.
\end{align}

Using the formulae (\ref{phiandH}), we can determine the vortex fields.
The Higgs field is 
\begin{align}
\phi(u) = \frac{\rmd f}{\rmd u} = \frac{2K}{\pi}
{\rm cn}\left(\frac{2K}{\pi}u \,; k \right)
{\rm dn}\left(\frac{2K}{\pi}u \,; k \right) \,,
\end{align}
and
\begin{align}
H = \frac{(\sin({\rm Re} \, u))^2}{\left({\rm Re} \, 
{\rm sn}\left(\frac{2K}{\pi}u \,; k \right)\right)^2} \,,
\end{align}
from which $|\phi|^2_H$ and other quantities are easily obtained.
Since the hyperbolic ends of $M$ are each mapped to the
imaginary axis in the $w$-plane, which is the boundary of $\HH^2$, the
solution satisfies the usual vortex boundary conditions.

This solution is a 2-vortex. Its two simple vortices are centred 
at opposite points along the geodesic circle 
${\rm Re} \, u = \frac{\pi}{2}$, at ${\rm Im} \, u = 0$ and  
${\rm Im} \, u = \frac{\lambda}{2}$.

Further solutions could be obtained by composing the map $f$ above
with maps from $\HH^2$ to itself, i.e. with suitable Blaschke functions.

\section{Interpretation in four dimensions}
\news

Generally, given a solution of the Bogomolny equations for abelian 
vortices on a hyperbolic surface $M$, one can construct a self-dual 
$SU(2)$ Yang-Mills field on the four-dimensional product manifold 
$M \times S^2$, where the sphere has the round metric with Gauss 
curvature $\half$. This Yang--Mills field is $SO(3)$-invariant over 
the $S^2$ factor. Explicit formulae for the Yang--Mills gauge field 
in complex coordinates are given by Popov \cite{Popov}.

In particular, as Witten showed \cite{Witten}, this construction 
gives a Yang--Mills gauge field on 
$\HH^2 \times S^2$ starting with a vortex solution on $\HH^2$, and
this in turn can give a Yang--Mills gauge field on $\R^4$, since the 
Yang--Mills equations are conformally invariant, and 
$\HH^2 \times S^2$ is conformal to $\R^4 - \R$, as one sees from the
manipulation of the metric 
\begin{align}
 \rmd s^2 &= \frac{2}{r^2}(\rmd\tau^2 + \rmd r^2) 
+ 2(\rmd\theta^2 + \sin^2\theta \, \rmd\varphi^2) \\
          &\cong \rmd\tau^2 + \rmd r^2 
+ r^2(\rmd\theta^2 + \sin^2\theta \, \rmd\varphi^2) \,.
\end{align}
Here, the metric on $\HH^2$ is the usual metric in the
UHP model, with $z = \tau + \rmi r$.
The excluded line $\R$ is the $\tau$-axis of $\R^4$, which corresponds
to the boundary of $\HH^2$, and is where the $SO(3)$-orbits collapse
from spheres to points. However, provided the vortex solution approaches
the vacuum on the boundary of $\HH^2$, then the gauge field can be extended
smoothly to this excluded line, resulting in a gauge field on
$\R^4$. If the vortex number is $\cal N$ then the $SU(2)$ gauge field on
$\R^4$ is a multi-instanton, with instanton number $\cal N$. These
$SO(3)$-invariant instantons can be interpreted as centred on the 
$\tau$-axis, even though the vortex centres are away from this axis.

This is the four-dimensional interpretation of the vortex examples 
discussed in subsection 4.1. 

All the hyperbolic surfaces that appear in subsections 4.2 -- 4.4 have
as universal covering space the hyperbolic plane, i.e. they are
quotients of the hyperbolic plane by a discrete infinite
group. This discrete group action commutes with the $SO(3)$ action on
$S^2$. Therefore the vortex solutions on these surfaces all lift to 
$\R^4 - \R$ as $SO(3)$-invariant, self-dual $SU(2)$ Yang--Mills gauge fields,
invariant also under the discrete group. The gauge fields have
infinite instanton number, and may not extend to the $\tau$-axis. Let us now
look at these examples in detail.

The compact surfaces of subsection 4.2 are quotients of $\HH^2$ 
by Fuchsian groups. Here the vortex fields, regarded as fields of
infinite vortex number on the covering space $\HH^2$, have no good limit as 
one approaches any point on the boundary of $\HH^2$. The lifted fields 
are therefore well defined on $\R^4 - \R$, but do not extend to the 
$\tau$-axis.  

The examples of subsection 4.3 are more interesting. Recall that
the hyperbolic surface here is the punctured disc, which is 
the hyperbolic plane quotiented by the translation group $\Z$, whose 
generator in the half-plane model is the translation 
$\tau \to \tau + 2\pi$. Lifted to $\R^4$, the fields are
still invariant under $\tau \to \tau + 2\pi$. Since the
vortex fields approach vacuum values on the $\tau$-axis (the
hyperbolic end of the punctured disc), they lift smoothly to all of 
$\R^4$. Quotienting by the translation group on $\R^4$ one 
obtains smooth $SU(2)$ gauge fields on $\R^3 \times S^1$, whose 
instanton number equals the initial vortex number. Such fields are 
called calorons. 

The vortex fields do not have vacuum boundary conditions at the
parabolic end (the puncture), where $r \to \infty$. However, they give the 
right asymptotic behaviour for a caloron. The caloron fields do not approach 
the standard Yang--Mills vacuum at infinity, as finite action
instantons on $\R^4$ would do.

Now, not all calorons have an $SO(3)$ symmetry, but certain calorons that
have been found explicitly have this symmetry, and have been described
in terms of a periodic holomorphic function on the hyperbolic plane,
i.e. a holomorphic function on the punctured disc \cite{HarShe,Harl}. 
These correspond to our vortex solutions (\ref{eq:cusp_sol}), with $f$
a Blaschke function as in (\ref{BlaschProd}).
It would be worthwhile to clarify the relation between the parameters of the
vortex solutions and the corresponding caloron fields and their 
boundary conditions.

A special case occurs if the discrete translation symmetry extends to
continuous translational symmetry in the $\tau$-direction. Then
the self-dual $SU(2)$ Yang-Mills field in $\R^4$ has no
$\tau$-dependence, and obeys the equation
\begin{align}
D_i A_{\tau} = -\half \epsilon_{ijk} F_{jk}
\end{align}
where $i,j,k$ run over the Cartesian indices of $\R^3$. This equation
is the Bogomolny equation for monopoles in $\R^3$, where the Higgs
field is identified as $A_{\tau}$. 

We can verify this relationship of hyperbolic vortices to $SU(2)$ monopoles
by evaluating the magnitude of the $SU(2)$ Higgs field (which is not
the Higgs field of the hyperbolic vortex, but rather the
component $-\half a_{\tau}$ of the vortex gauge potential). We work
in the punctured disc picture, with the complex coordinate 
$u= \rme^{\rmi z}= \rme^{\rmi\tau}e^{-r}$. To have
continuous translation invariance in $\R^4$, we must select the function $f(u)$
to be circularly symmetric, i.e. $f(u) = u^k$ with $k$ a positive
integer. Then the vortex Higgs field and the bundle metric are
\begin{align}
 \phi(u) = k u^{k-1} \quad {\rm and} \quad
 H(u, \bar u) = \frac{u\bar{u} (\log u\bar{u})^2}
{\left(1 - (u\bar{u})^k \right)^2}. 
\end{align}
This Higgs field is not $\tau$-independent, but the $\tau$-dependence 
can be removed by the gauge transformation $g = u^{-k+1}$, after which
\begin{align}
 &\phi \to g\phi = k \,, \\
 &H \to g^{-1}{\bar g}^{-1}H 
 = \frac{(u\bar{u})^k (\log u\bar{u})^2}
   {\left(1 - (u\bar{u})^k \right)^2} \,. \label{monopoleH} 
\end{align}
The gauge transformation $g$ is admissible on $D^*$ as the origin is 
not included, and since $g$ is holomorphic, one does not leave 
holomorphic gauge so the connection is still the Chern connection,
\begin{align}
a_u = \rmi\pr_u(\log H) \,, \quad a_{\bar u} = 0 \,.
\end{align}
To obtain the $SU(2)$ Higgs field, note that
\begin{align}
 -\half\left(a_{\tau} - \rmi a_r\right) = -a_z = -\rmi\pr_z(\log H) 
= u\pr_u(\log H) \,,
\end{align}
and from the expression (\ref{monopoleH}) for $H$ we find that 
$a_r = 0$ and
\begin{align}
 -\half a_{\tau} = k\coth kr - \frac{1}{r} \,. 
\end{align}
This is precisely the magnitude of the Higgs field 
of a spherically symmetric monopole. It is unsurprising that the 
monopole charge, the coefficient of the $\frac{1}{r}$ term, is 1, 
since no higher charge $SU(2)$ monopoles are spherically symmetric. 
Since the asymptotic magnitude of the Higgs field is $k$, the mass of 
the monopole is $4\pi k$. Lifted to $\R^3 \times S^1$,
the Yang--Mills action is $8\pi^2 k$ so the instanton number is $k$.   

Finally, we briefly discuss the interpretation of the vortex solutions
on the hyperbolic cylinder, constructed in subsection 4.4. The
cylinder is obtained from the hyperbolic plane by quotienting by the 
discrete dilation group generated by $(\tau, r) \to \rme^{-\lambda} (\tau,
  r)$. The group lifts to $\R^4$ as a group of dilations there, so the
lifted gauge fields are defined on the quotient of $\R^4$ by this
group, which is the manifold $S^3 \times S^1$, where
$S^3$ parametrises angles in $\R^4$ and $S^1$ is a finite radial interval
with ends identified. Because the vortex fields take vacuum values at 
both ends of the hyperbolic cylinder, no singularities occur on the
circles over the two points of $S^3$ which intersect the $\tau$-axis. 
Our vortex construction in terms of the Jacobi function gives 
explicit instantons on 
$S^3 \times S^1$. These have $SO(3)$ symmetry, with $SO(3)$ acting
on $S^3$ in the same way that the diagonal subgroup of $SU(2) \times SU(2)$ 
acts on $SU(2)$.

A general study of instantons on $S^3 \times S^1$ was made by Braam
and Hurtubise \cite{BraHur}, and our solutions are a subset of these 
instantons.

\news 

\section*{Acknowledgements}

NSM is grateful to Ian Strachan, Nigel Hitchin, and especially 
Duong Hong Phong for illuminating discussions and guidance.
NAR thanks EPSRC, the Cambridge European Trust, and St.~John's
College, Cambridge for financial support.

\end{document}